\documentclass[preprint,12pt]{aastex}
\usepackage{epstopdf}

\newcommand{\kms}{km~s$^{-1}$}

\newcommand{\cmms}{cm$^2$~s$^{-1}$}
\newcommand{\msun}{M$_{\sun}$}
\newcommand{\rjup}{R$_{Jup}$}
\newcommand{\ldl}{$\lambda/{\Delta}{\lambda}$}

\newcommand{\teff}{T$_{eff}$}
\newcommand{\logg}{$\log{g}$}

\newcommand{\kzz}{$K_{zz}$}

\newcommand{\meth}{CH$_4$}

\newcommand{\wat}{H$_2$O}
\newcommand{\ammon}{NH$_3$}
\newcommand{\name}{ULAS~J141623.94+134836.3}
\newcommand{\namesh}{ULAS~J1416+1348}

\slugcomment{Submitted to ApJ 2 Feb 2010}

\shorttitle{The Blue T Dwarf ULAS~J1416+1348}
\shortauthors{Burgasser, Looper \& Rayner}

\begin{document}

\title{ULAS~J141623.94+134836.3: a Blue T Dwarf Companion to a Blue L Dwarf}

\author{
Adam J.\ Burgasser\altaffilmark{1,2},
Dagny Looper\altaffilmark{3,4}
and John T.\ Rayner\altaffilmark{3,4}
}

\altaffiltext{1}{Center for Astrophysics and Space Science, University of California San Diego, La Jolla, CA 92093, USA; aburgasser@ucsd.edu}
\altaffiltext{2}{Massachusetts Institute of Technology, Kavli Institute for Astrophysics and Space Research, Building 37, Room 664B, 77 Massachusetts Avenue, Cambridge, MA 02139, USA}
\altaffiltext{3}{Institute for Astronomy, University of Hawaii, 2680 Woodlawn Drive, Honolulu, HI 96822, USA}
\altaffiltext{4}{Visiting Astronomer at the Infrared Telescope Facility, which is operated by
the University of Hawai`i under Cooperative Agreement NCC 5-538 with the National Aeronautics
and Space Administration, Office of Space Science, Planetary Astronomy Program}

\begin{abstract}
We confirm the substellar nature of ULAS~J141623.94+134836.3, a common proper motion companion
to the blue L dwarf SDSS~J141624.08+134826.7 identified by Burningham et al.\ and Scholz. 
Low-resolution 0.8--2.4~$\micron$ spectroscopy obtained with IRTF/SpeX shows
strong {\wat} and {\meth} absorption bands, consistent with a T7.5 spectral type, and we see possible indications of {\ammon} absorption in the 1.0--1.3~$\micron$ region.  More importantly,
the spectrum of ULAS~J1416+1348 shows a broadened $Y$-band peak and highly suppressed $K$-band flux, both indicative of high surface gravity and/or subsolar metallicity.  These traits are verified through spectral model fits, from which we derive atmospheric parameters {\teff} = 650$\pm$60~K, {\logg} = 5.2$\pm$0.4~cgs, [M/H] $\leq$ -0.3 and {\kzz} = 10$^4$~{\cmms}, the temperature being significantly warmer than that estimated by Burningham et al.  These fits also indicate a model-dependent spectroscopic distance of 10.6$^{+3.0}_{-2.8}$~pc for ULAS~J1416+1348, formally consistent with the 7.9$\pm$1.7~pc astrometric distance for SDSS~J1416+1348 from Scholz.  
The common peculiarities of these two co-spatial, co-moving sources suggest that their unusual blue colors---and those of other blue L and T dwarfs in general---arise from
age or metallicity, rather than cloud properties alone.  
\end{abstract}


\section{Introduction}

Over the past 15 years, well over 500 brown dwarf members of the late-M, L and T dwarf spectral classes have been identified in various Galactic environments, encompassing a broad diversity in color, spectral properties, and physical characteristics (\citealt{2005ARA&A..43..195K} and references therein).  Sustained effort has been made to identify the coldest of these sources, which include the low-mass extreme of star formation and primordial relics of the Galactic halo population.  The most recent discoveries made with the Two Micron All Sky Survey (2MASS; \citealt{2003yCat.2246....0C,2006AJ....131.1163S}), the UKIRT Infrared Deep Sky Survey (UKIDSS; \citealt{2007MNRAS.379.1599L}) and the Canada France Hawaii Telescope Legacy Survey (CFHTLS; \citealt{2008A&A...484..469D}) have extended the known population down to and below effective temperatures {\teff} $\approx$ 600~K
(e.g., \citealt{2007MNRAS.381.1400W,2008MNRAS.391..320B, 2009MNRAS.395.1237B, 2008ApJ...689L..53B,2008A&A...482..961D,2009ApJ...695.1517L}).  This has raised the question as to where the currently coldest class of brown dwarfs---the T dwarfs---ends and the next cooler class---the Y dwarfs---might begin.  Such exceedingly dim and cold sources are predicted to encompass several major chemical transitions in brown dwarf atmospheres, including the disappearance of Na and K into salt condensates, the emergence of strong {\ammon} absorption across the infrared band, and the formation of photospheric water ice clouds (e.g., \citealt{1999ApJ...513..879M, 1999ApJ...519..793L, 2002Icar..155..393L, 2003ApJ...596..587B}).  Accordingly, there is considerable interest and controversy as to how to delineate this putative class; 
see discussions by \citet{2007ApJ...667..537L,2008MNRAS.391..320B} and \citet{2008A&A...482..961D}.  

A promising low-temperature brown dwarf candidate was recently identified by \citet{2010arXiv1001.4393B} and \citet{2010arXiv1001.2743S} 
as a co-moving companion to
the nearby blue L6 dwarf SDSS~J141624.08+134826.7 (hereafter SDSS~J1416+1348; \citealt{2010ApJ...710...45B, 2009arXiv0912.3565S, kirkpatrick2010}).  The object, {\name} (hereafter {\namesh}), was identified in UKIDSS as a faint ($J$ = 17.35$\pm$0.02) and unusually blue ($J-K = -1.58~\pm$~0.17) near-infrared source separated by 9$\farcs$8 from the L dwarf.  Using astrometry from 2MASS, 
UKIDSS, the Sloan Digital Sky Survey Data Release 7 (SDSS DR7;  \citealt{2000AJ....120.1579Y,2009ApJS..182..543A}), and follow-up imaging, both \citet{2010arXiv1001.4393B} and
\citet{2010arXiv1001.2743S} were able to confirm common proper motion of this pair.  \citet{2010arXiv1001.2743S} also determined an astrometric distance to the primary of 7.9$\pm$1.7~pc, consistent with spectrophotometric estimates from \citet[6.5--10.7~pc]{2010ApJ...710...45B} and \citet[6.4--9.6~pc]{2009arXiv0912.3565S}.  At this distance, the (poorly constrained) absolute magnitudes of
{\namesh}, $M_J$ =  17.8$\pm$0.5 and $M_K$ = 19.4$\pm$0.5, are
equivalent to or fainter than those of the latest-type brown dwarfs with measured distances, Wolf 940B 
($M_J$ =  17.68$\pm$0.28 and $M_K$ = 18.37$\pm$0.28; \citealt{2009MNRAS.395.1237B}) and ULAS~J003402.77$-$005206.7 
($M_J$ =  17.65$\pm$0.11 and $M_K$ = 17.98$\pm$0.12; \citealt{2007MNRAS.381.1400W,2009arXiv0912.3163S}).  
\citet{2010arXiv1001.4393B} report a 1.0--2.5~$\micron$ spectrum of {\namesh}, identifying it
as a T7.5 brown dwarf with highly suppressed $K$-band flux.
Indeed, {\namesh} is the bluest T dwarf in $J-K$ color identified to date, matching the unusually blue nature
of its L dwarf companion.
{\it Spitzer} photometry reported in \citet{2010arXiv1001.4393B} further 
suggest an exceptionally low-temperature ({\teff} $\sim$ 500~K), metal-poor ([M/H] $\approx$ -0.3) and high surface gravity atmosphere ({\logg} $\approx$ 5.0--5.3~cgs).

In this article, we report our measurement of the near-infrared spectrum of {\namesh} obtained with the NASA Infrared Telescope Facility (IRTF) SpeX spectrograph \citep{2003PASP..115..362R}.  This spectrum encompasses the 0.8--2.4~$\micron$ region, including the metallicity-sensitive $Y$-band peak.  The unusual shape of this and the $K$-band flux peak, along with fits to spectral models, affirm the interpretation of this source as a metal-poor, high surface gravity T7.5 brown dwarf, albeit with a {\teff} that is significantly warmer than that reported
by \citet{2010arXiv1001.4393B}.    
In Section~2 we describe our observations and discuss the spectral properties of {\namesh}, including its classification, spectral anomalies and possible indications of {\ammon} absorption in the 1.0--1.3~$\micron$ region.   In Section~3 we present our spectral model fits to the data and corresponding atmospheric parameters, as well as a model-dependent spectroscopic distance that is in accord with the astrometric distance of the primary.  We discuss the relevance of this system with regard to the nature of blue L and T dwarfs in Section~4.  Results are summarized in Section~5.

\section{Near Infrared Spectroscopy}

\subsection{Observations and Data Reduction}

Low resolution near-infrared spectral data for {\namesh} were
obtained with SpeX on 2010 January 23 (UT) in mostly clear skies with some light cirrus and 
0$\farcs$8 seeing.  We used the SpeX prism mode with the 0$\farcs$5 slit aligned to the parallactic angle
for all observations, providing 0.7--2.5~$\micron$
coverage in a single order with resolution R~$\equiv$ {\ldl} $\approx 120$
and dispersion of 20--30~{\AA}~pixel$^{-1}$.
{\namesh} was acquired with the slit viewing camera using the $J$ filter
and guiding was performed on the nearby primary.
A total of 34 exposures of 180~s each were obtained in ABBA dither pairs,
nodding along the slit.  The first 16 exposures of the source were obtained over
an airmass range of 1.28--1.41.  We then observed the A0~V star HD~121880 ($V$ = 7.59) 
at an airmass of 1.12 for flux calibration and telluric absorption correction, as well as 
internal flat field and argon arc lamps for pixel response and wavelength calibration.
{\namesh} was then reacquired and 18 more exposures made over an airmass range of 1.05--1.13.

Data were reduced with the IDL SpeXtool package, version 3.4
\citep{2004PASP..116..362C}, using standard settings.
Due to the faintness of {\namesh} and its highly structured spectral morphology, individual spectra
were optimally extracted using a trace of HD~121880 as a template.  These
spectra were then combined using a robust weighted average after scaling each to the median flux at the $J$-band peak. Telluric absorption and instrumental response corrections were determined from the A0~V spectrum following the method of \citet{2003PASP..115..389V}, 
with line-shape kernels derived from the arc lines and adjustments made to the H I line strengths and wavelength scale, as outlined in \citet{2004PASP..116..362C}.  

\subsection{The Spectrum of {\namesh}}

The reduced spectrum of {\namesh} is shown in Figure~\ref{fig_nirspec}, 
compared to equivalent data for the T8 dwarfs 2MASS~J04151954$-$0935066 (hereafter 2MASS~J0415$-$0935; \citealt{2002ApJ...564..421B, 2004AJ....127.2856B})
and 2MASS~J09393548$-$2448279 (hereafter 2MASS~J0939$-$2448; \citealt{2005AJ....130.2326T, 2006ApJ...637.1067B}).  
{\namesh} exhibits the unambiguous signatures of a T dwarf, with
strong {\wat} and {\meth} absorption and a blue spectral energy distribution.  The 1.6~$\micron$ {\meth} band in the spectrum of {\namesh} is slightly weaker than those of the T8 comparison sources, although the breadth of the $J$- and $H$-band peaks (both shaped by the wings of {\wat} and {\meth} bands) are equivalent 
to the spectrum of 2MASS~J0415$-$0935.   T dwarf classification indices \citep{2006ApJ...637.1067B} indicate a spectral type of T7.5$\pm$0.5 for this source, which is also consistent with its {\ammon}-H and $W_J$ indices (Table~\ref{tab_indices};  \citealt{2007MNRAS.381.1400W,2008MNRAS.391..320B, 2008A&A...482..961D}).  This classification and most of the spectral indices are in agreement with those determined by \citet{2010arXiv1001.4393B}.  However, we find a significant disagreement in our measurement of the {\meth}-J index.\citet{2010arXiv1001.4393B} specifically note this index as anomalousm whereas our value is consistent with the overall spectral classification of {\namesh}.  As both spectra were obtained at roughly the same resolution ({\ldl} $\approx$ 100), and signal-to-noise of the SpeX data in the $Y$-, $J$- and $H$-band peaks is good ($\sim$40--70), the origin of this anomaly is unclear.

What is most remarkable about the spectrum of {\namesh} is the breadth of its 1.07~$\micron$ $Y$-band flux peak and strongly suppressed 2.10~$\micron$ $K$-band flux peak.  The latter feature was also noted in the spectral data of \citet{2010arXiv1001.4393B}, and explains the very blue $J-K$ color of the source; we calculate a spectrophotometric color of $J-K = -1.71\pm0.23$\footnote{This value was determined by calculating the colors of 100 realizations of the spectral data, with fluxes varied following a normal distribution of the noise spectrum.  We report here the mean and standard deviation of these measures.} from our SpeX spectrum, consistent with both the UKIDSS photometry and measurements by \citet{2010arXiv1001.4393B}.   The broadened $Y$-band peak in the spectrum of {\namesh}
is readily apparent, and similar to but more extreme than the broadened peak seen in the spectrum of 2MASS~J0939$-$2448 (Figure~\ref{fig_nirspec}; see also Figure~2 in \citealt{2006ApJ...639.1095B}).  The origins of both features are discussed below.

\subsection{{\ammon} Absorption in the 1.0--1.3~$\micron$ Region?}

In addition to these broad spectral anomalies, we identified several intriguing absorption features around the $Y$-, $J$- and $H$-band flux peaks in the spectrum of {\namesh}. As shown in Figure~\ref{fig_nirfeatures}, these features are at 0.997, 1.039, 1.072, 1.232, 1.249, 1.292, 1.302 and 1.570~$\micron$, none of which are present in the late-type T dwarfs with comparable SpeX data.\footnote{See \url{http://www.browndwarfs.org/spexprism}.}  Among these features, the 1.072, 1.232 and 1.570~$\micron$ features are also seen in the absorption spectrum of Jupiter \citep{2009ApJS..185..289R}.  Given tentative suggestions of the onset of {\ammon} absorption in the near-infrared spectra of  the latest-type T dwarfs \citep{2000ApJ...541..374S, 2007ApJ...667..537L,2008A&A...482..961D}, we examined whether any of these features might be coincident with {\ammon} opacity.  Figure~\ref{fig_nirfeatures} overlays the laboratory transmission function of {\ammon} from \citet{1999JQSRT..62..193I}, measured at temperatures of 200--300~K and pressures of 0.01--1~bar.  Structure in the 
{\ammon} spectrum appears to be coincident with 
some of the features, most notably the 0.997, 1.039 and prominent 1.072~$\micron$ dips in the $Y$-band, and the weaker 1.292 and 1.302~$\micron$ dips in the $J$-band.  However, strong {\ammon} opacity features, such as the 1.01--1.05 and 1.19--1.23~$\micron$ bands, are not seen in the data.

There are important caveats to such comparisons of opacity measurements to low-resolution brown dwarf spectra.  First, opacity from several species, most notably {\wat} and {\meth} gas, blankets the entire near-infrared region, and the resolution of the SpeX data makes it impossible to separate narrow features from these species from coincident absorption arising from to {\ammon}.  Second, the laboratory measurements of \citet{1999JQSRT..62..193I} were obtained in very different gas conditions than those that characterize the warmer photospheres of T dwarfs, and are not likely to include the higher angular momentum states expected in to be present in brown dwarf spectra.   Indeed, \citet{2007ApJ...667..537L} have shown that current brown dwarf models incorporating the \citet{1999JQSRT..62..193I} opacities predict {\ammon} bands that are much stronger than observed, even when nonequilibrium abundances due to vertical mixing are considered \citep{2006ApJ...647..552S, 2007ApJ...669.1248H}.  

In summary, while these features are notable, they cannot be conclusively associated with {\ammon} absorption.  Higher resolution spectra coupled with better opacity data are needed to verify their origin.

\section{The Physical Properties of {\namesh}}

\subsection{Qualitative indicators of High Surface Gravity and Subsolar Metallicity}

The broadened $Y$-band and suppressed $K$-band peaks in the spectrum of {\namesh} are similar in nature to those seen in previously identified, unusually blue T dwarfs, and are indicative of pressure effects related to surface gravity and metallicity \citep{2002ApJ...564..421B, 2006ApJ...639.1095B, 2004AJ....127.3553K,2006AJ....131.2722C, 2007ApJ...667..537L, 2009ApJ...702..154S}.  
$K$-band flux is regulated by collision-induced H$_2$ absorption \citep{1969ApJ...156..989L, 1994ApJ...424..333S, 2002A&A...390..779B}, which is sensitive to both photospheric gas temperature and pressure.
 The short wavelength slope of the $Y$-band peak is shaped by the red wing of the pressure-broadened 0.77~$\micron$ \ion{K}{1} doublet, which is also modulated by temperature (affecting the K abundance) and pressure (affecting the pressure-broadened wings; \citealt{2003A&A...411L.473A,2003ApJ...583..985B}). 
 The $Y$-band peak has been specifically noted as being metallicity-sensitive in comparison of synthetic atmosphere
 models, becoming both broadened and blue-shifted for lower metallicities \citep{2006ApJ...639.1095B, 2007ApJ...667..537L}.  

The archetype blue T dwarf, 2MASS~J09373487+2931409 (hereafter 2MASS~J0937+2931; \citealt{2002ApJ...564..421B}) exhibits the same $Y$-band and $K$-band anomalies as {\namesh} and, importantly, is consistently well-matched to models with subsolar metallicities ([M/H] = $-$0.1 to $-$0.4) and high surface gravities ({\logg} = 5.2 to 5.5; \citealt{2006ApJ...639.1095B,2009ApJ...695..844G}).  
2MASS~J0939$-$2448 also exhibits these peculiarities (Figure~\ref{fig_nirspec}), and its near- and mid-infrared spectrum is well-matched to subsolar metallicity models as well, although it is additionally suspected of being an unresolved binary \citep{2008ApJ...689L..53B,2009ApJ...695.1517L}.   
Importantly, the $Y$-band and $K$-band anomalies are more pronounced in the spectrum of {\namesh} than in those of 2MASS~J0937+2931 and 2MASS~J0939$-$2448.
Our measure of the $K/J$ index---the relative flux between the $J$- and $K$-band peaks---is the smallest reported to date: 0.037$\pm$0.004 compared to 0.059 for 2MASS~J0939$-$2448 (see Table~6 in \citealt{2006ApJ...637.1067B} and Table~6 in \citealt{2009MNRAS.395.1237B}).
These measures suggest that {\namesh} is a true outlier in terms of its physical properties.

\subsection{Comparison to Spectral Models}

To quantify these properties, we compared the spectrum of {\namesh} to the 
atmosphere models of \citet{2008ApJ...689.1327S}.  We followed the prescriptions detailed in \citet{2008ApJ...678.1372C} and \citet{2008ApJ...689L..53B}, comparing our SpeX spectrum, flux-calibrated to the $J$-band photometry reported in \citet{2010arXiv1001.4393B}, to models spanning temperatures
{\teff} = 500--1000~K (50~K steps); surface gravities {\logg} = 4.0--5.5~cgs (0.5~dex steps); metallicities
[M/H] = $-$0.3, 0 and +0.3 dex relative to Solar; and vertical diffusion coefficients {\kzz} = 0 and 10$^4$~{\cmms} (see \citealt{2006ApJ...647..552S}).  The models were smoothed to the resolution of the SpeX data
using a Gaussian kernel, and interpolated onto a common wavelength scale.
Fits were made exclusively to the 0.9--2.4~$\micron$ region.  The goodness-of-fit statistic $G_k$ \citep{2008ApJ...678.1372C} was used to gauge the agreement between models and data, and we followed the same weighting scheme employed by those authors in which each pixel is weighted by its breadth in wavelength space.  Model surface fluxes were scaled by the factor $C_k = (R/d)^2$ which minimizes $G_k$ (Equation~2 in \citealt{2008ApJ...678.1372C}), where $R$ is the radius of the brown dwarf and $d$ its distance from the Earth.  
Fits were made to all 264 models.  Distributions of the fit parameters were generated following a weighting scheme similar to that described in \citet{2008ApJ...689L..53B}, in which each model's parameters were incorporated into the distributions with a weight proportional to\footnote{Note that in \citet{2008ApJ...689L..53B}, the weighting function was  e$^{-0.1G_k}$, a conservative choice that favored poorer-fitting models more highly.  The 0.5 coefficient used here is more consistent with the probability distribution function of the $\chi^2$ statistic, for which $G_k$ is a close analog.} e$^{-0.5G_k}$.  To examine the robustness of our fits to observational uncertainties, we also performed a Monte Carlo simulation similar to that described in \citet{2008ApJ...678.1372C} and \citet{2009ApJ...706.1114B}, generating 1000 realizations of the spectrum with fluxes randomly varied about the measured values following a normal distribution of the observational noise; the overall scaling of the spectrum was also varied following a normal distribution tied to the uncertainty in the $J$-band photometry.  These spectra were compared to the 20 models that best fit the original spectrum, and distributions of the resulting best-fit parameter sets and $C_k$ scale factors were determined. 

Table~\ref{tab_fits} summarizes the parameters of the ten best-fitting spectral models, while Figure~\ref{fig_modelfit} shows the overall best-fit model overlaid on the spectrum of {\namesh}: {\teff} = 650~K, {\logg} = 5.0~cgs, [M/H] = $-$0.3 and {\kzz} = 10$^4$~{\cmms}.  This model was the best fit for all of the spectra in our Monte Carlo simulation; i.e., its Monte Carlo fraction $f_{MC}$ = 1.000 \citep{2008ApJ...678.1372C}.  It is a reasonably good match to the spectral
energy distribution, qualitatively reproducing the strong absorption bands, broadened $Y$-band peak and suppressed $K$-band peak, although the $H$-band peak flux is $\sim$10-15\% underestimated.  
The parameter distributions from all of the model fits are also shown in Figure~\ref{fig_modelfit}.  
Gaussian fits to these distributions yield optimal parameters of {\teff} = 650$\pm$60~K and {\logg} = 5.2$\pm$0.4~cgs.  
The metallicity distribution clearly favors a metal-poor atmosphere; the five best-fitting models all have a subsolar metallicity.  In fact, the one-sided distribution in our limited model set means that we 
cannot rule out metallicities less than $-0.3$.   The model fits also indicate some vertical mixing is present, favoring {\kzz} = 10$^4$~{\cmms} over 0~{\cmms}, although a strict constraint cannot be made.

With respect to surface gravity and metallicity, our fits to the spectrum of {\namesh} are in agreement with
the estimates of \citet{2010arXiv1001.4393B}, indicating that this unusually blue T dwarf is likely to be old, massive and metal-poor.  The derived {\teff} and {\logg} parameters and their uncertainties correspond to
an age of 2--10~Gyr and a mass of 0.021--0.045~{\msun} according to the evolutionary models of \citet{2003A&A...402..701B}.  This age is consistent with membership in the Galactic disk population, as previously suggested by the system's kinematics \citep{2010ApJ...710...45B, 2009arXiv0912.3565S}.  The subsolar metallicity favored by the model fits is in quantitative agreement with spectral analyses of other blue T dwarfs, as well as characterization of the L dwarf companion, SDSS~J1416+1348, which does not appear to be a full-fledged L subdwarf (\citealt{2010ApJ...710...45B}; however, see \citealt{kirkpatrick2010}).   As such, these fits support our qualitative analysis of the spectrum: the spectral peculiarities and blue color of {\namesh} appear to be the result of a high pressure atmosphere arising from high surface gravity and subsolar metallicity.

\subsection{The {\teff} of {\namesh}}

Our inferred {\teff} for {\namesh} is somewhat low for T7--T8 dwarfs, which typically have {\teff} $\approx$ 700--900~K \citep{2004AJ....127.3516G,2004AJ....127.2948V, 2009ApJ...702..154S}.  This is likely to be a surface gravity and/or metallicity effect.
\citet{2006ApJ...639.1095B} have previously found that late-type T dwarfs with higher surface gravities tend to have lower {\teff}s for a given spectral type.  
\citet{2010arXiv1001.4393B}, on the other hand, derive an even lower temperature for {\namesh}, {\teff} $\approx$ 500~K, based on this source's uniquely red $H$-[4.5] color.   The link between {\teff} and $H$-[4.5] color for brown dwarfs cooler than $\sim$1000~K was originally established by \citet{2007MNRAS.381.1400W}, and has been shown to provide increased sensitivity for the latest-type T dwarfs \citep{2009ApJ...702..154S,2010arXiv1001.0762L}.  However, \citet{2009ApJ...695.1517L} have noted that metallicity effects are relevant and can shift $H$-[4.5] to the red by roughly 0.1~mag for every 0.1~dex decrement in metallicity (see also Figure~6 in \citealt{2010arXiv1001.4393B}).  If {\namesh} has a metallicity significantly below [M/H] = $-$0.3---not ruled out by the present model fits---then this characteristic may have as much to do with its extreme color as its low temperature.
It is relevant to note that the [3.6]-[4.5] color of {\namesh}, another {\teff} indicator \citep{2006ApJ...651..502P}, is not an extremum; this source is in fact bluer than 2MASS~J0939$-$2448.  This may be an indication that the 3.3~$\micron$ {\meth} band, like the 1.6~$\micron$ band, is relatively weak compared to other T8-T9 dwarfs, consistent with a warmer {\teff}.  However, metallicity and/or surface gravity effects may again complicate a strict correlation.

The \citet{2008ApJ...689.1327S} spectral model based on the atmosphere parameters favored by \citet{2010arXiv1001.4393B}---{\teff} = 500~K, {\logg} = 5.0~cgs, [M/H] = $-$0.3~dex and {\kzz} = 10$^4$~{\cmms---is also shown in Figure~\ref{fig_modelfit}.  The cooler model actually provides a better match to the relative flux between the $J$- and $H$-band peaks and the width of the $J$-band peak; but predicts stronger 1.6~$\micron$ {\meth} absorption, a far more distorted $Y$-band peak and a more suppressed $K$-band flux peak than observed.  These deviations make this model a 2$\sigma$ outlier compared to the best-fit model for our data.  
We emphasize that the differences between these fits do not explicitly rule out either set of parameters It is well known that incomplete opacity tables, inaccurate treatment of \ion{K}{1} pressure broadening and the influence of distributed condensate opacity (``cloud tops'') can result in poor fits to T dwarf near-infrared spectra \citep{2006ApJ...639.1095B, 2007ApJ...656.1136S, 2008ApJ...678.1372C,2009ApJ...702..154S}.  However, to the limits of the accuracy of the current spectral models, our analysis favors a warmer temperature for {\namesh} than indicated by its $H$-[4.5] color.

\subsection{Spectroscopic Distance}

Following \citet{2009ApJ...706.1114B}, we calculated a spectroscopic parallax for {\namesh} using the model-to-data flux scaling factor $C_k$ derived from the spectral modeling.  The mean value and uncertainty of this factor (based on the same $G_k$ weighting scheme used for the parameter distributions) yields a distance-to-radius ratio d/R = 12.8$\pm$3.0~pc/{\rjup}.  Based on the inferred {\teff} and {\logg} range, the evolutionary models of \citet{2008ApJ...689.1327S} predict a radius $R$ = 0.83$^{+0.14}_{-0.10}$~{\rjup}, corresponding to a distance of 10.6$^{+3.0}_{-2.8}$~pc.   This is larger than but within 1$\sigma$ of the astrometric distance of the primary from \citet{2010arXiv1001.2743S}, 7.9$\pm$1.7~pc.  In contrast, the 500~K model shown in Figure~\ref{fig_modelfit} requires d/R = 5.8~pc/{\rjup}, and the corresponding $R$ = 0.73~{\rjup} implies a distance of only 4.2~pc, significantly smaller than both spectrophotometric and astrometric estimates for SDSS~J1416+1348.  Hence, to the limits of the accuracy of the spectral and evolutionary models of \citet{2008ApJ...689.1327S}, our atmospheric parameter determinations for {\namesh} are commensurate with this source being cospatial with its co-moving L dwarf companion.

\section{The Nature of Blue L and T Dwarfs}

The SDSS~J1416+1348/{\namesh} system provides a unique opportunity to explore the underlying physical properties that distinguish blue L and T dwarfs.  While surface gravity and metallicity effects have long 
been acknowledged as contributors to the peculiarities of blue T dwarfs, 
condensate cloud properties have been seen as playing a more important role in
shaping the spectra of blue L dwarfs.  Several studies have argued that thin and/or patchy condensate clouds in the photospheres of blue L dwarfs adequately explain their unique photometric and spectroscopic characteristics \citep{2004AJ....127.3553K, 2007AJ....133..439C,2008ApJ...674..451B, 2009ApJ...702..154S}.  However, thin clouds cannot be responsible for the colors and spectra of late-type blue T dwarfs---such as {\namesh}---since clouds are buried deep below the visible photosphere in these low-temperature objects  \citep{2001ApJ...556..872A}.  

The distinct empirical characteristics shared by SDSS~J1416+1348 and {\namesh} must have an origin that is common to both sources; this argues for age and/or metallicity.  Older ages for blue L and T dwarfs are supported by their collective kinematics; \citet{2009AJ....137....1F} and \citet{kirkpatrick2010} have shown that this subgroup exhibits
a much broader range of tangential velocities ($\sigma_{V} \approx$ 50~{\kms}) than L and T dwarfs with ``normal'' colors ($\sigma_{V} \approx$ 22~{\kms}).  The high surface gravities inferred from spectral model fits to blue L and T dwarfs further support older ages for these sources (e.g., \citealt{2008ApJ...689L..53B, 2008ApJ...678.1372C, 2009ApJ...695..844G}).  Subsolar metallicities are also supported by spectral model fits to blue T dwarfs, and the fact that blue L dwarfs exhibit spectral characteristics that are intermediate between normal field L dwarfs and halo L subdwarfs \citep{2004ApJ...614L..73B,kirkpatrick2010}.  However, the discovery of a blue L5 dwarf companion to the solar-metallicity field M4.5 star G~203-50 \citep{2008ApJ...689..471R} suggests that metallicity does not play a consistent role in shaping these spectra.  

We argue that the common photometric and spectroscopic properties of SDSS~J1416+1348 and {\namesh} favors old age, and possibly subsolar metallicity, as the physical trait that characterizes the blue L and T dwarf populations.
Thin condensate clouds may still be common for blue L dwarf atmospheres, with higher surface gravities and subsolar metallicities contributing to increased sedimentation rates and a reduced supply of condensate species, respectively.  
However, our conjecture predicts that these cloud properties are simply a consequence of the high-pressure photospheres characterizing old,  high surface gravity and---in some cases---metal-poor brown dwarfs.

\section{Summary}

We have measured the 0.8--2.4~$\micron$ spectrum of {\namesh}, the common proper motion companion
to the blue L dwarf SDSS~J1416+1348.
These data confirm the T7.5 spectral type determined by \citet{2010arXiv1001.4393B},
show possible {\ammon} features in the 1.0--1.3~$\micron$ region,
and reveal broadened $Y$-band and highly suppressed $K$-band peaks consistent with a
high surface gravity and/or subsolar metallicity.  Spectral model fits based on calculations by
\citet{2008ApJ...689.1327S} indicate atmospheric
parameters {\teff} = 650$\pm$60~K, {\logg} = 5.2$\pm$0.4~cgs, [M/H] $\leq$ -0.3 and {\kzz} = 10$^4$~{\cmms}.  The metallicity and surface gravity are consistent with the analysis by \citet{2010arXiv1001.4393B}, but our {\teff} is $\sim$150~K (2.5$\sigma$) warmer.  If correct, it suggests that the extreme $H$-[4.5] color of this source
may be due to metallicity and/or surface gravity effects, rather than an exceedingly low {\teff}.  Our fit parameters for {\namesh} imply a model-dependent spectroscopic distance that is formally consistent with the astrometric distance of SDSS~J1416+1348 measured by \citet{2010arXiv1001.2743S}, and further strengthens the case that this pair is a coeval system of
unusually blue brown dwarfs.   We argue that the common peculiarities of the SDSS~J1416+1348/{\namesh} system implies that most unusually blue L and T dwarfs derive their unique properties from old age, and possibly subsolar metallicity, with the thin clouds of blue L dwarfs being a secondary effect.   

Despite the substantial amount of follow-up already done for this fairly recent discovery, its benchmark role in understanding temperature, surface gravity, metallicity and cloud effects in L and T dwarf spectra motivates further observational study of both components.  These include independent parallax measurements to verify absolute fluxes; higher-resolution near-infrared spectroscopy and mid-infrared spectroscopy of the secondary to validate potential {\ammon} features and discern the origin of its unusual mid-infrared colors; broad-band spectral energy distribution measurements of both components to measure luminosities and constrain {\teff}s; high-resolution imaging to search for additional components; and improved model fits to better constrain atmospheric parameters.
In addition, the $\sim$100~AU projected separation of this system---wider than any L dwarf/T dwarf pair identified to date---raises questions as to the formation of it and other widely-separated, very low-mass stellar/brown dwarf multiples (e.g., \citealt{2004ApJ...614..398L,2005A&A...440L..55B, 2007ApJ...660.1492C}).  Coupled with its proximity to the Sun,  the SDSS~J1416+1348/{\namesh} system is a target of opportunity
for studies of cold brown dwarf atmospheres and origins.

\acknowledgements

The authors acknowledge telescope operator Dave Griep 
at IRTF for his assistance with the observations; 
B.\ Burningham for providing an electronic version of the Irwin et al.\ {\ammon} opacity spectrum,
and D.\ Saumon for providing electronic copies of the spectral and evolutionary models used in the analysis.
This research has benefitted from the M, L, and T dwarf compendium housed at DwarfArchives.org and maintained by Chris Gelino, Davy Kirkpatrick, and Adam Burgasser; 
the SpeX Prism Spectral Libraries, maintained by Adam Burgasser at 
\url{http://www.browndwarfs.org/spexprism}; and
the VLM Binaries Archive maintained by Nick Siegler at \url{http://www.vlmbinaries.org}.
The authors recognize and acknowledge the 
very significant cultural role and reverence that 
the summit of Mauna Kea has always had within the 
indigenous Hawaiian community.  We are most fortunate 
to have the opportunity to conduct observations from this mountain.

Facilities: \facility{IRTF~(SpeX)}

\clearpage

\begin{figure}
\epsscale{0.8}
\plotone{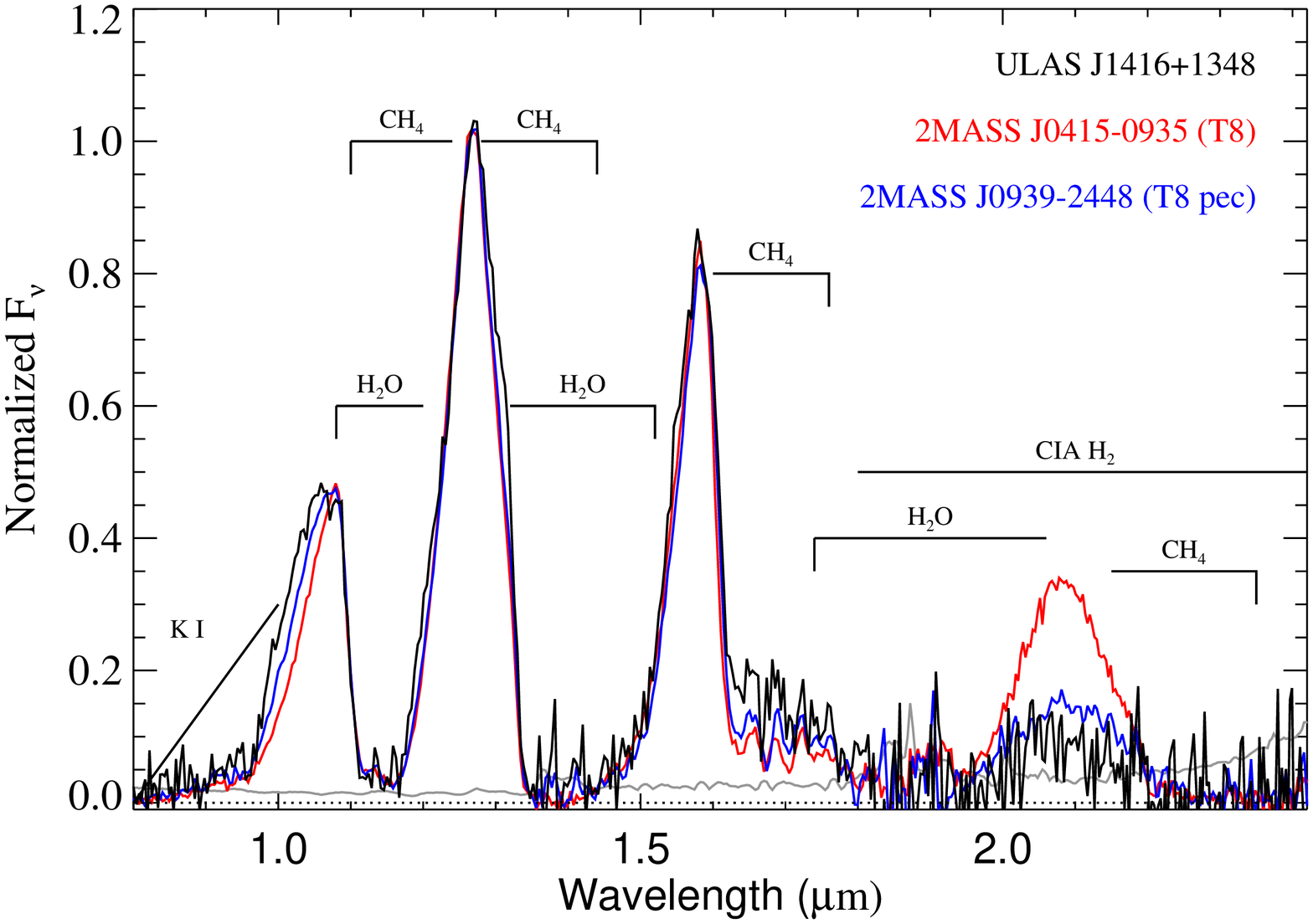}
\caption{SpeX prism spectrum of {\namesh} (black line) compared
to equivalent data for the T8 dwarfs
2MASS~J0415$-$0935 (red line; \citealt{2004AJ....127.2856B})
and 2MASS~J0939$-$2448 (blue line; \citealt{2006ApJ...637.1067B}).
All three spectra
are normalized at 1.27~$\micron$, and the corresponding noise spectrum for {\namesh}
is indicated by the light grey line.  
Prominent {\wat} and {\meth} absorption features are labeled, as well as the region influenced by the pressure-broadened \ion{K}{1} doublet wing ($\lambda \lesssim 1$~$\micron$) and collision-induced H$_2$ opacity ($\lambda \gtrsim 1.75$~$\micron$).
\label{fig_nirspec}}
\end{figure}

\clearpage

\begin{figure}
\epsscale{1.0}
\plotone{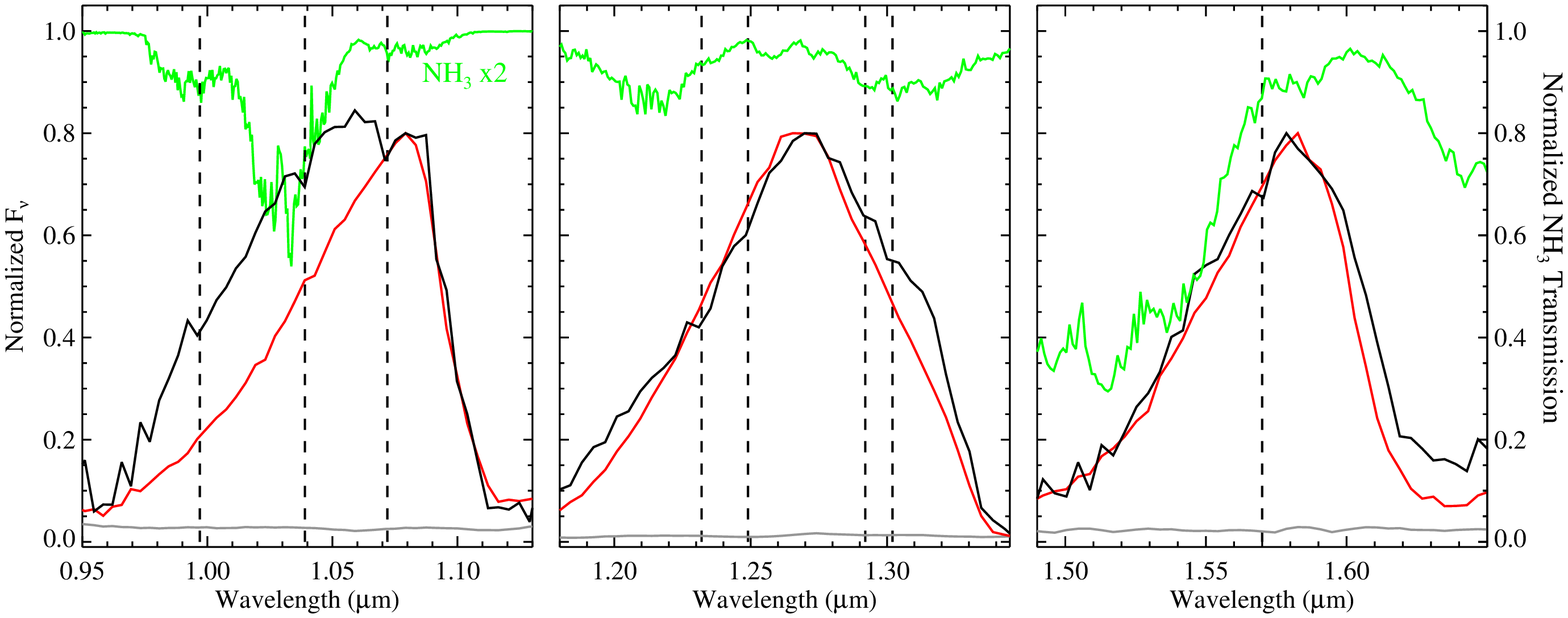}
\caption{Close-up views of the 1.07~$\micron$ ($Y$-band, left), 1.27~$\micron$ ($J$-band, middle) and 1.58~$\micron$ ($H$-band, right) flux peaks in the spectra of {\namesh} (black line) and
2MASS~J0415$-$0935 (red line).  
Data are normalized in each panel to the peak flux in the given band.
Also shown is the normalized transmission spectrum of {\ammon}
from \citet{1999JQSRT..62..193I}.  The transmission is magnified by a factor of two in the $Y$-band panel to highlight weaker bands.
The absorption features in the spectrum of {\namesh} noted in the text are indicated by dashed lines.
\label{fig_nirfeatures}}
\end{figure}

\clearpage

\begin{figure}
\epsscale{1.0}
\includegraphics[width=0.95\textwidth]{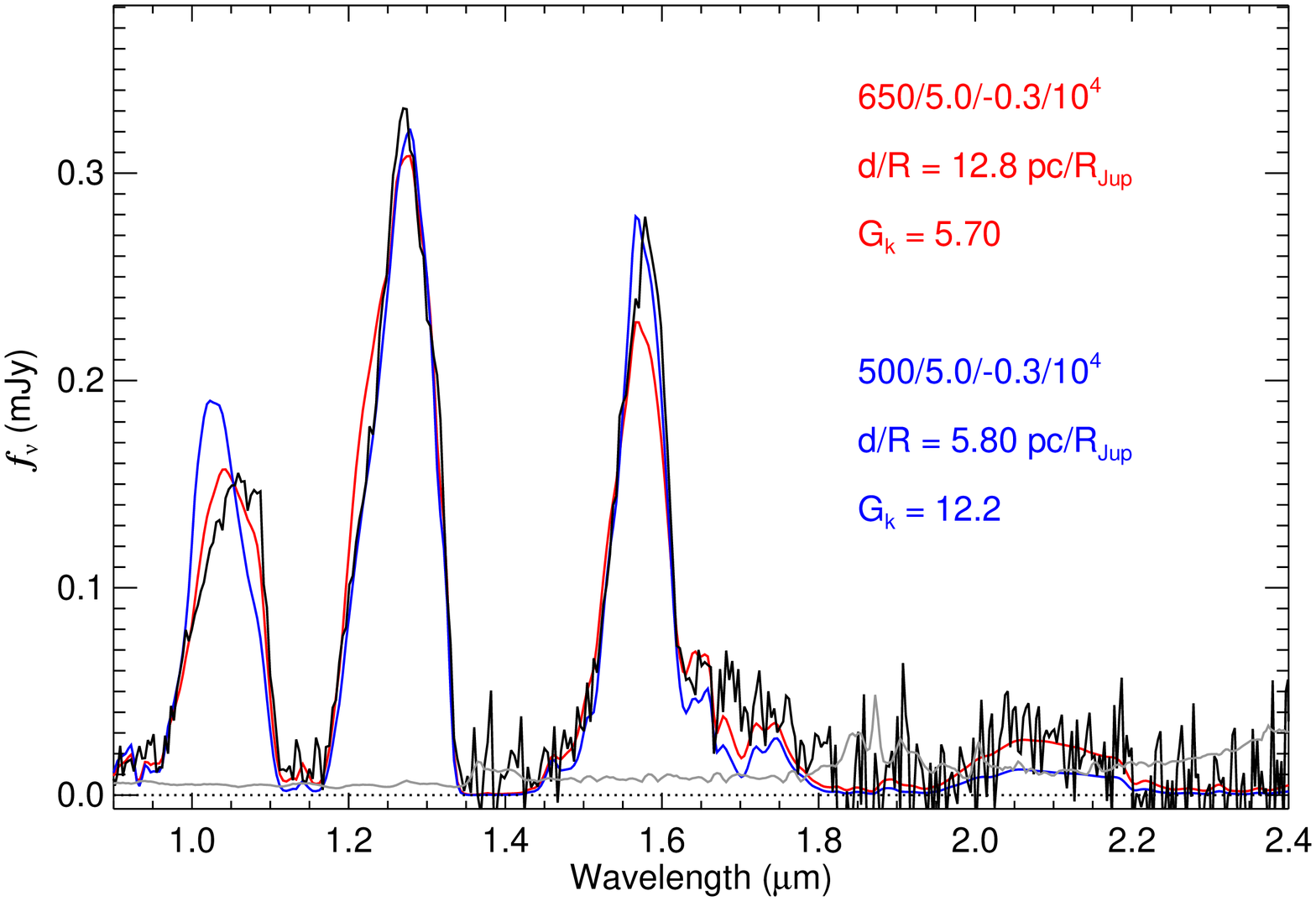}
\includegraphics[width=0.24\textwidth]{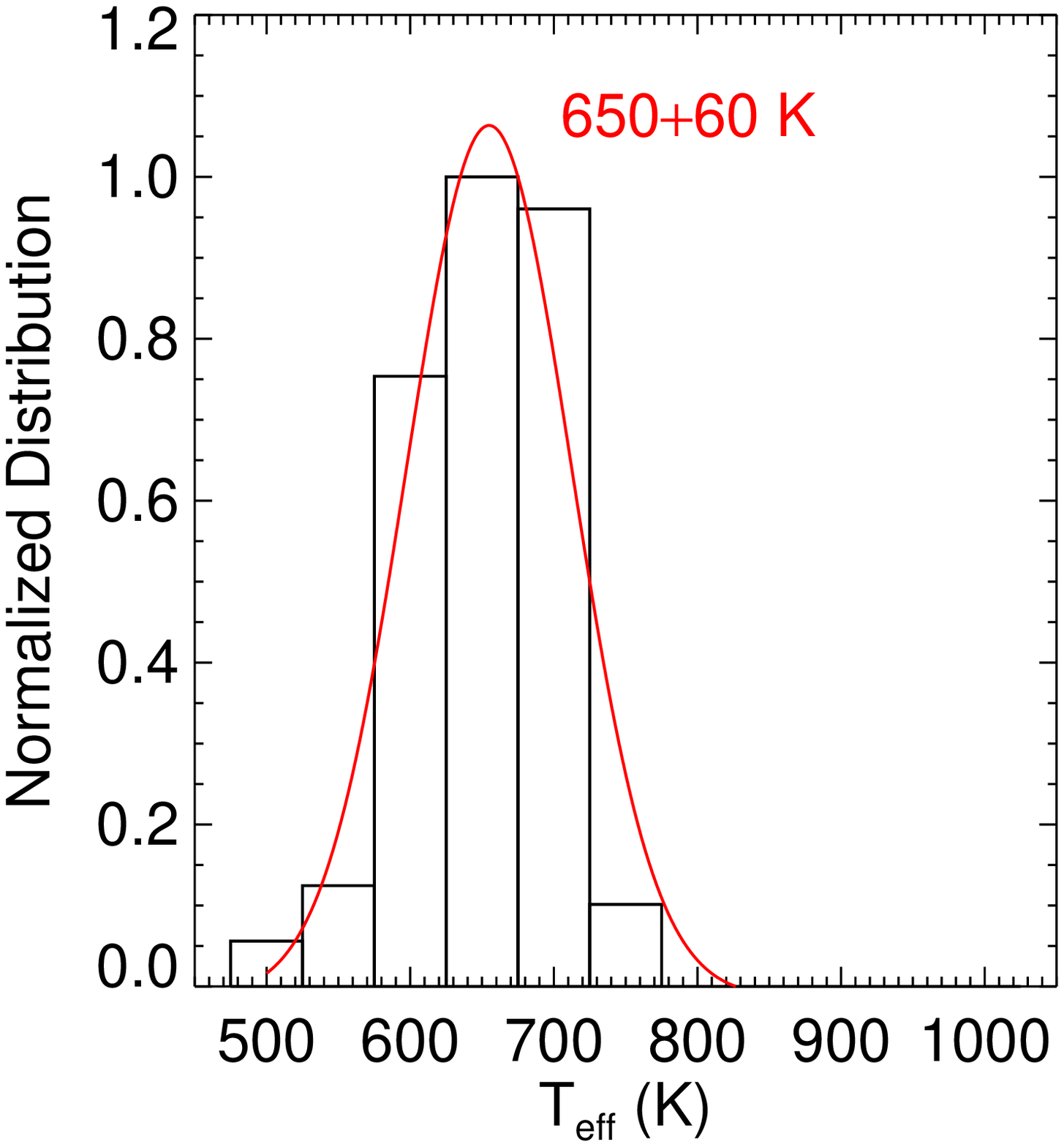}
\includegraphics[width=0.24\textwidth]{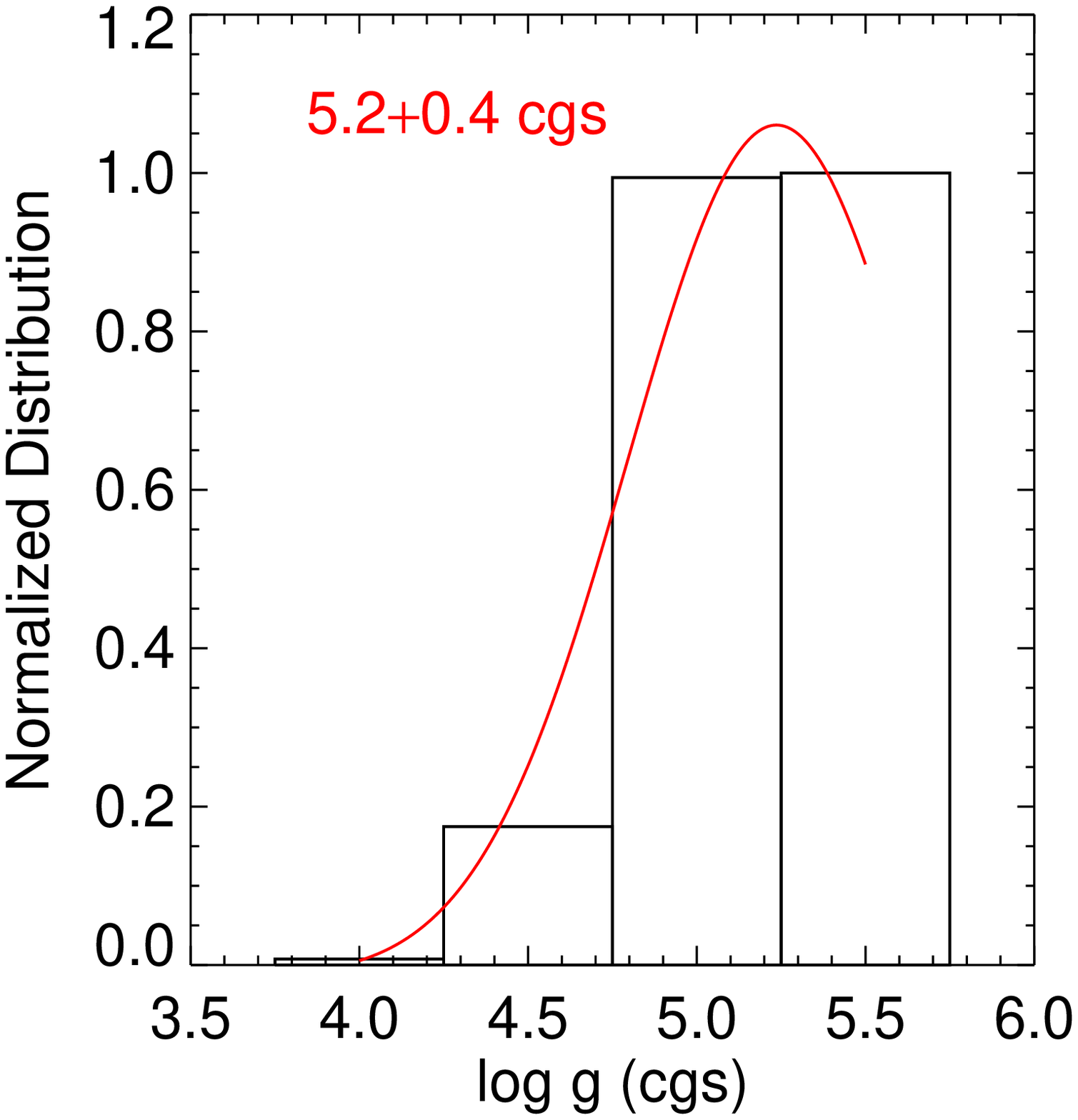}
\includegraphics[width=0.24\textwidth]{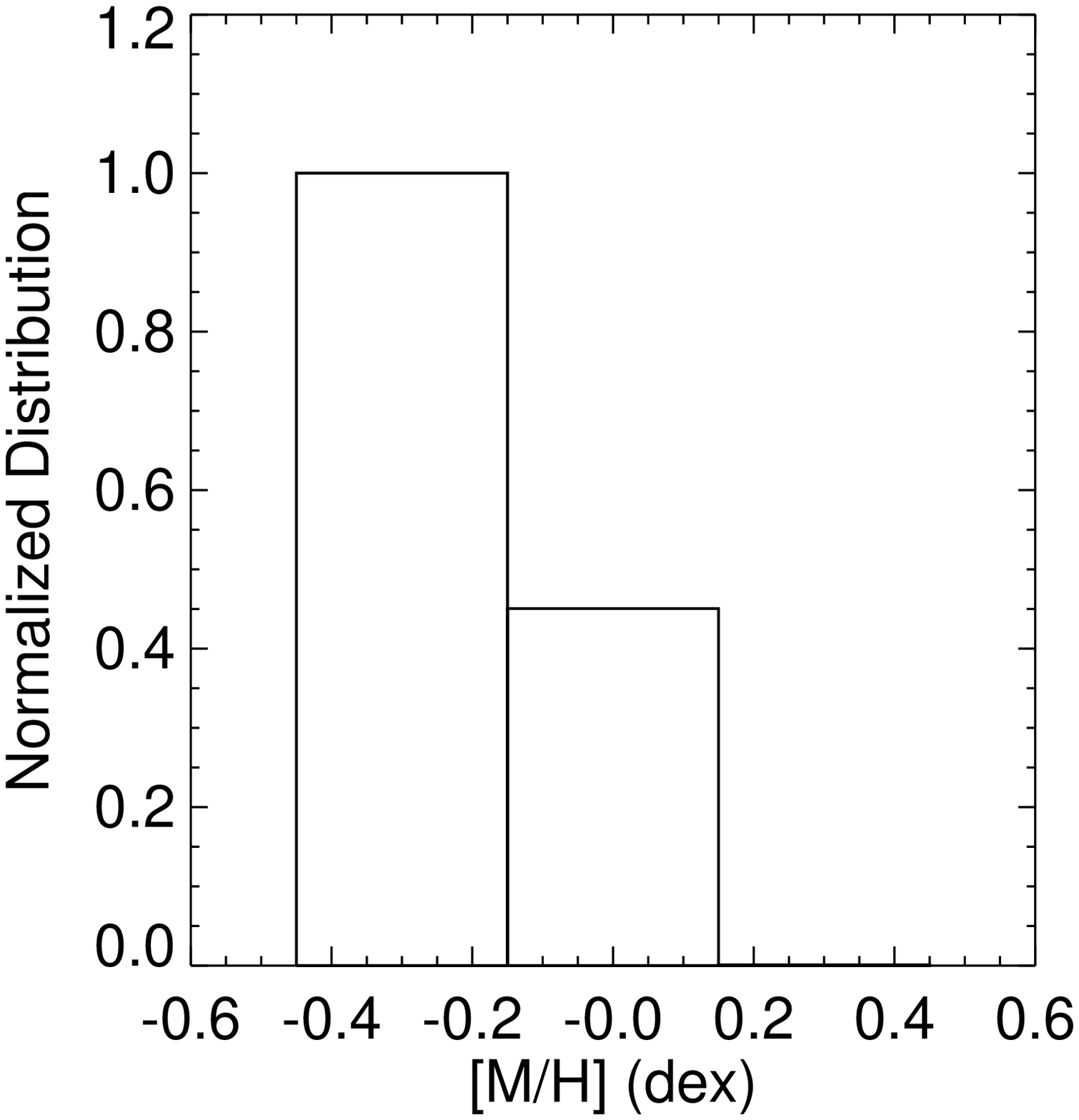}
\includegraphics[width=0.24\textwidth]{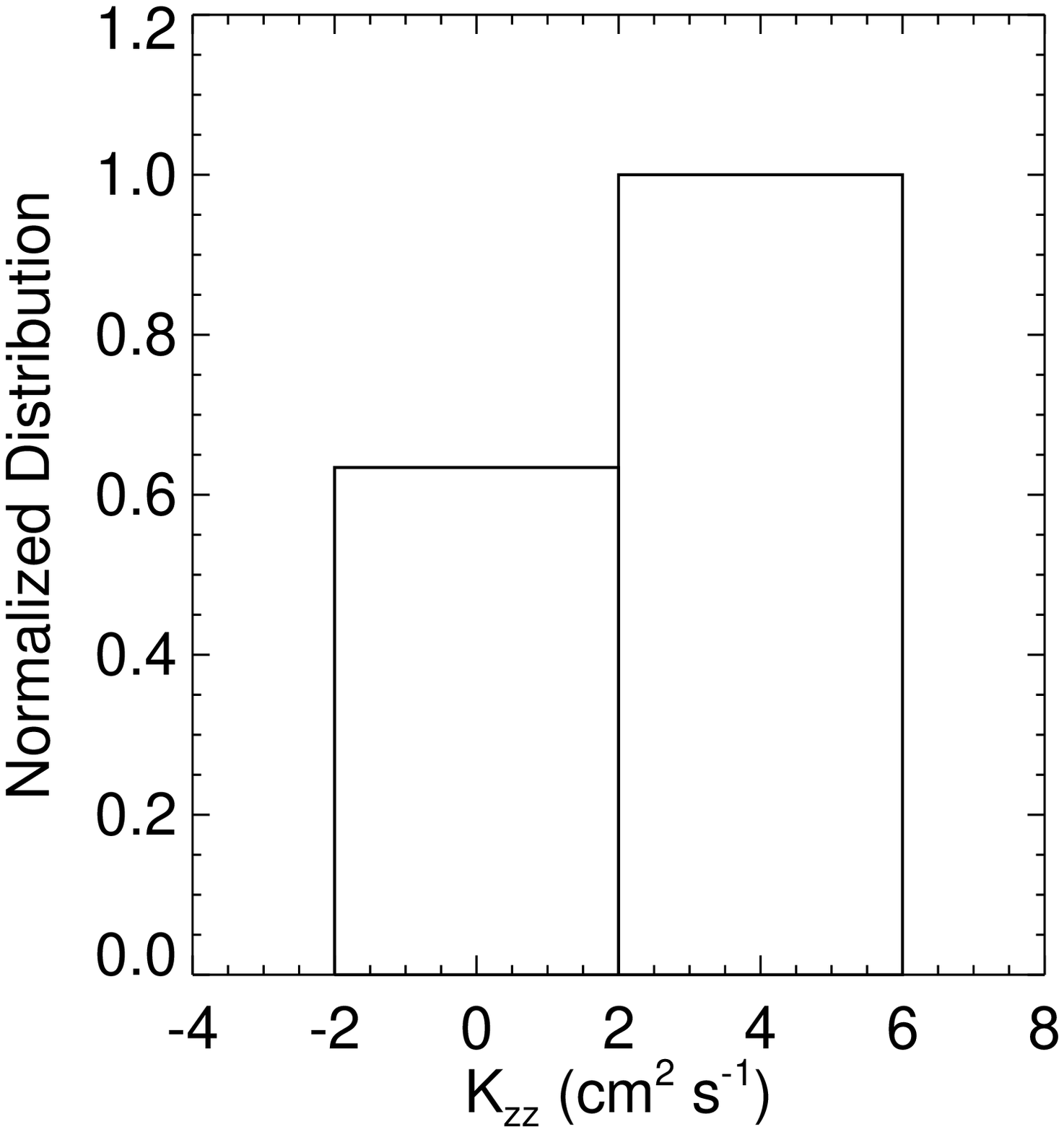}
\caption{(Top panel): Best-fitting model spectrum (red line) to our SpeX spectrum of {\namesh} (black line): {\teff} = 650~K, {\logg} = 5.0~cgs, [M/H] = -0.3 and {\kzz} = 10$^4$~{\cmms}; and the model corresponding to the parameters of \citet{2010arXiv1001.4393B}: {\teff} = 500~K, {\logg} = 5.0~cgs, [M/H] = -0.3 and {\kzz} = 10$^4$~{\cmms}.  The data are scaled to $J$-band photometry from \citet{2010arXiv1001.4393B} and the models are scaled to minimize $G_k$ values (indicated). 
The noise spectrum for {\namesh} is indicated by the grey line.
(Bottom panels): Parameter distributions of (left to right) {\teff}, {\logg}, [M/H] and {\kzz} based on the weighting scheme described \citet{2008ApJ...689L..53B} and in the text.   Means and uncertainties for {\teff} and {\logg} are indicated in the first two panels and are based on Gaussian fits to the distributions.  The metallicity distribution is such that we can only conclude that [M/H] $\leq$ -0.3, while the {\kzz} distribution indicates a slight preference for {\kzz}  = 10$^4$~{\cmms}.
 \label{fig_modelfit}}
\end{figure}

\clearpage

\begin{deluxetable}{ccccc}
\tabletypesize{\footnotesize}
\tablecaption{Spectral Indices for {\namesh}. \label{tab_indices}}
\tablewidth{0pt}
\tablehead{
\colhead{Index} &
\colhead{Value\tablenotemark{a}} &
\colhead{SpT} &
\colhead{Value B10\tablenotemark{b}} &
\colhead{Reference}  \\
}
\startdata
{\wat}-J & 0.053$\pm$0.008 & T8 &  0.07$\pm$0.01 & 1  \\
{\meth}-J & 0.268$\pm$0.006 & T7 &  0.34$\pm$0.01 &  1  \\
$W_J$ & 0.376$\pm$0.005 & T7  &   0.34$\pm$0.01 & 2,3  \\
{\wat}-H & 0.181$\pm$0.011 & T8 &   0.20$\pm$0.01 & 1  \\
{\meth}-H & 0.197$\pm$0.010 & T7 &   0.20$\pm$0.01 & 1  \\
{\ammon}-H & 0.675$\pm$0.014 & \nodata  &  0.61$\pm$0.01 &  4  \\
{\meth}-K & 0.085$\pm$0.144 & T7 &  0.29$\pm$0.02 &  1  \\
K/J & 0.037$\pm$0.004 & \nodata &  \nodata &  1  \\
\enddata
\tablenotetext{a}{Spectral index values were measured for 1000 realizations of the spectrum, each with a normal distribution of random values scaled by the noise spectrum added to the original fluxes.  The reported values are the means and standard deviations of these measurements.}
\tablenotetext{b}{Spectral index values reported in \citet{2010arXiv1001.4393B} based on {\ldl} $\approx$ 100 near-infrared spectral data.}
\tablerefs{(1) \citet{2006ApJ...637.1067B}; (2) \citet{2007MNRAS.381.1400W}; (3) \citet{2008MNRAS.391..320B}; (4) \citet{2008A&A...484..469D}.}
\end{deluxetable}

\begin{deluxetable}{lcccccc}
\tabletypesize{\footnotesize}
\tablecaption{Ten Best-Fitting \citet{2008ApJ...689.1327S} Spectral Models to SpeX Data for {\namesh}. \label{tab_fits}}
\tablewidth{0pt}
\tablehead{
\colhead{Rank} &
\colhead{\teff~(K)} &
\colhead{\logg~(cgs)} &
\colhead{[M/H]~(dex)} &
\colhead{\kzz~({\cmms})} &
\colhead{$G_k$} &
\colhead{$d/R$ (pc/{\rjup})} 
 \\
}
\startdata
1\tablenotemark{a} & 650 & 5.0 & -0.3 & 10$^4$ & 5.70 & 12.8 \\
2 & 700 & 5.0 & -0.3 & 10$^4$ & 5.97 & 15.5 \\
3 & 600 & 5.0 & -0.3 & 10$^4$ & 6.10 & 10.2 \\
4 & 700 & 5.0 & -0.3 & 0 & 6.53 & 15.4 \\
5 & 650 & 5.0 & -0.3 & 0 & 6.58 & 12.7 \\
6 & 600 & 5.5 & 0.0 & 10$^4$ & 6.85 & 10.5 \\
7 & 650 & 5.5 & 0.0 & 10$^4$ & 6.88 & 13.3 \\
8 & 700 & 5.5 & 0.0 & 10$^4$ & 7.31 & 15.7 \\
9 & 600 & 5.0 & -0.3 & 0 & 7.37 & 10.2 \\
10 & 650 & 5.5 & 0.0 & 0 & 8.01 & 13.3 \\
\cline{1-7}
Avg.\tablenotemark{b} & 650$\pm$60 & 5.2$\pm$0.4 & $\leq$-0.3 & $\sim10^4$ & \nodata & 12.8$\pm$3.0 \\
\enddata
\tablenotetext{a}{Best-fitting model for 1000 synthesized spectra in Monte Carlo simulation; i.e., $f_{MC} = 1.000$ (see \citealt{2008ApJ...678.1372C}).}
\tablenotetext{b}{Based on the weighted parameter distributions shown in Figure~\ref{fig_modelfit}.  Each model contributes its parameters to the distributions scaled by the factor $e^{-0.5G_k}$. The means and uncertainties of {\teff} and {\logg} were determined by Gaussian fits to their respective distributions (see \citealt{2008ApJ...689L..53B}).  The [M/H] distribution peaked at the lower limit of the sampled parameter space, while the models slightly favor {\kzz} = 10$^4$~{\cmms} over 0~{\cmms}.}
\end{deluxetable}

\clearpage


\end{document}